# SOFFLFM: Super-resolution optical fluctuation Fourier light-field microscopy

**Haixin Huang,**[1] **Haoyuan Qiu,**[1] **Hanzhe Wu,**[1] **Yihong Ji,**[1] **Heng Li,**[1,2] **Bin Yu,**[1] **Danni Chen**[1,*] **and Junle Qu**[1]

[1] *College of Physics and Optoelectronic Engineering, Key Laboratory of Optoelectronic Devices and Systems of Ministry of Education and Guangdong Province, Shenzhen University, Shenzhen 518060, China*

[2] *Tsinghua-Berkeley Shenzhen Institute (TBSI), Tsinghua University, Shenzhen, 518055, China*

*\* danny@szu.edu.cn*

**Abstract:** Fourier light-field microscopy (FLFM) uses a micro-lens array (MLA) to segment the Fourier Plane of the microscopic objective lens to generate multiple two-dimensional perspective views, thereby reconstructing the three-dimensional(3D) structure of the sample using 3D deconvolution calculation without scanning. However, the resolution of FLFM is still limited by diffraction, and furthermore, dependent on the aperture division. In order to improve its resolution, a Super-resolution optical fluctuation Fourier light field microscopy (SOFFLFM) was proposed here, in which the Sofi method with ability of super-resolution was introduced into FLFM. SOFFLFM uses higher-order cumulants statistical analysis on an image sequence collected by FLFM, and then carries out 3D deconvolution calculation to reconstruct the 3D structure of the sample. Theoretical basis of SOFFLFM on improving resolution was explained and then verified with simulations. Simulation results demonstrated that SOFFLFM improved lateral and axial resolution by more than $\sqrt{2}$ and 2 times in the 2$^{nd}$ and 4$^{th}$ order accumulations, compared with that of FLFM.

## 1. Introduction

In order to understand the basic principles of biological systems, it is essential to observe the three-dimensional (3D) structure of intracellular organelles with high spatial and temporal resolution. In traditional fluorescence microscopies, 3D information of samples is always collected in a sequential or scanning way. This process inevitably increases photo damage to living cells and reduces temporal resolution. Light-field microscopy (LFM) technology can simultaneously record two-dimensional spatial and angle information of light, allowing the 3D structure of a sample to be reconstructed from a single shot of the sample without any scanning.[1] The high scalability and high temporal resolution of LFM makes it useful in both functional brain imaging and single-cell imaging.[2-4] However, the inevitable reconstruction artifacts and huge computational costs limit the application of LFM.[5]

Recently, Fourier light-field microscopy (FLFM) has been developed to obtain four-dimensional light field information in the Fourier domain, so that the point spread function (PSF) of the system can be described by a unified 3D PSF.[6] This method effectively removes the limitation of LFM because of its reconstruction artifacts and computational cost, and furthermore the image quality is improved compared with the traditional LFM, which makes it more applicable in some fields such as endoscope imaging.[7]

However, it should be noted that, the resolution of FLFM is directly related to the aperture division of the system.[6] With an aperture partition coefficient $N$, the resolution is $N$ times worse than the diffraction-limited resolution. Recently, FLFM systems have recently been demonstrated to achieve resolution close to the diffraction limit in three dimensions with appropriate aperture division.[8] In fact, $N$ is numerically greater than 1, which enables FLFM to obtain multiple 2D perspective views (Angle information) to extract 3D information. Only by adjusting aperture division, the resolution that FLFM can achieve is inevitably lower than that in wide-field microscope. In order to solve this problem, an effective strategy was proposed, which replaced the on-axis central perspective collected by FLFM with high-resolution image acquired by wide-field microscopy.[9] This method effectively improved the lateral resolution of the system. However, because the off-axis perspective images which determines the axial resolution of FLFM system remained unchanged, the axial resolution was not improved. More recently, a few super-resolved FLFM were proposed. FLFM combined with single-molecule localization was verified to achieve 3D super-resolution imaging, which can achieve a localization accuracy of 20 nm in three dimensions within a depth range of 3 μm,[10] but it needs to collect tens of thousands of raw images for localization and reconstruction.

Here, we propose a 3D super-resolved FLFM, named Super-resolution optical fluctuation Fourier light-field microscopy (SOFFLFM). As an approach combining Super-resolution optical fluctuation imaging (SOFI) and FLFM, SOFFLFM can improve the resolution of FLFM in three dimensions and even break through the diffraction limitation by calculating high-order cumulants.

## 2. Theory of SOFFLFM

As an approach combining FLFM and Sofi, SOFFLFM could use the same imaging system as is used in previous FLFM.[6]

The light field propagation model of original FLFM can be described as shown in Fig. 1(a). Firstly, the 3D information of the object's imaging domain is mapped into the native imaging plane through Debye diffraction theory of the high numerical objective lens.[11] Then, the optical Fourier transform is performed through a Fourier lens on the native imaging plane, which means the spectral plane of the objective lens is relayed to the rear focal plane of the Fourier lens. In the rear focal plane of the Fourier lens, a micro-lens array (MLA) performs aperture segmentation and modulation on the frequency information collected by the objective lens. Finally, the light field from MLA is propagated to the rear focal plane to form the perspective images on the sensor. Obviously, there are two kinds of effects derived by spectral information segmentation and modulation through a MLA. First, with the division of aperture, the frequency information collected by the objective lens **is equalized by the two-dimensional perspective images produced by each sub-lens in MLA**. The lateral resolution of the FLFM system can be expressed as $R_{xy} = \frac{\lambda}{2NA_{FLFM}} = \left(\frac{\lambda}{2NA_{obj}}\right)N$, where $\lambda$ is the emission wavelength, $NA_{FLFM}$ is the equivalent numerical aperture of the FLFM system, and $NA_{obj}$ is the numerical apertures of the objective.[6] Therefore, compared with the traditional wide field microscopic system, it has lower resolution. Second, each sub-lens captures its local wavefront on the spectral plane, and **its position and direction on the rear focal plane are proportional to the average gradient of the captured wavefront**.[10] Therefore, the images of emitters at different axial positions on each perspective image are transversely translated to different positions through each independent sub-lens. FLFM system uses these lateral offsets to retrieve the axial information through 3D deconvolution. It is reasonable to take the images of emitters at different axial positions as Gaussian spots with different transversal offsets, and the axial resolution of the FLFM system is directly dependent on its lateral resolution.

Here, the FLFM's 3D PSF could be approximately described as:

$$U_{FLFM}(r) = U_{FLFM}(x,y,z) = \sum_j^{mla} exp(-\frac{(x-x_{zj})^2+(y-y_{zj})^2}{2\omega_0^2}) , \qquad (1)$$

where $(x_{zj}, y_{zj})$ is the lateral offset relative to the image center for the $j^{th}$ sub-lens. $\omega_0$ is the standard deviation of the Gaussian distribution.

The basic idea of SOFI is to improve the resolution of the whole image by calculating high-order cumulants of a time image sequence of the emitters with random fluctuations.[12] Due to the independence of fluctuations, signals from different emitters in 3D space are not correlated. Therefore, here in SOFFLFM, high-order cumulants calculation is carried out on a time series of perspective images collected by the above FLFM system to improve the resolution of each perspective images. Based on cumulants formula in Ref. 12 and Eq. (1), we give the $n^{th}$-order cumulant formula of FLFM as:

$$C_n(r,\tau_1,...,\tau_{n-1}) = \sum_k [U_{FLFM}^n(r) * \delta(r\text{-}r_k)]\varepsilon_k^n w_k(\tau_1,...,\tau_{n-1}) , \qquad (2)$$

where $\delta(r\text{-}r_k)$ is the distribution of emitters, $\varepsilon_k$ is the constant molecular brightness, $\tau$ is the time lag, $w_k(\tau_1,...,\tau_{n-1})$ is a correlation-based weighting function, depending on the specific fluctuation properties of each emitter in 3D space.

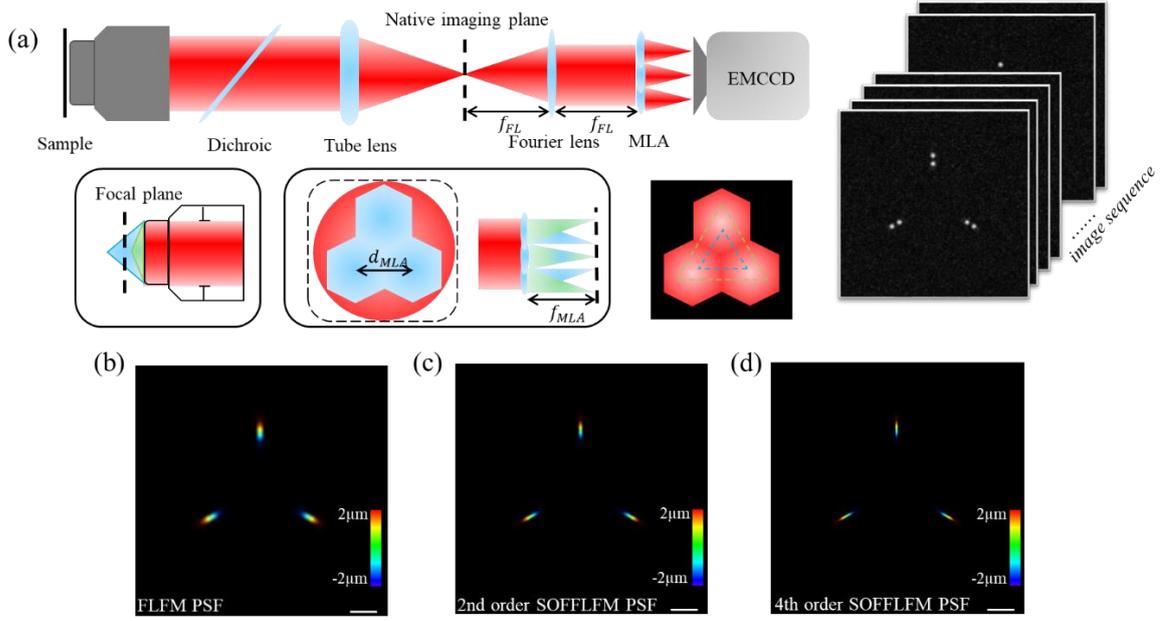

Fig. 1.   Schematic setup of SOFFLFM (a) and PSFs of the FLFM (b), $2^{nd}$ (c) and $4^{th}$ (d) order cumulants, with depth color-coded. Scale bar: 1 μm.

The value of the $n^{th}$-order cumulants defines the $n^{th}$-order SOFFLFM raw image whose PSF is the $n^{th}$ power of the original PSF. Therefore, the resolution of the $n^{th}$-order SOFFLFM raw image consisting of perspective images can theoretically be improved by $\sqrt{n}$. Finally, the 3D volume of the sample is reconstructed by deconvolving the $n^{th}$-order raw Sofi image with the $n^{th}$-order cumulative PSF, based on Richardson-Lucy algorithm.

There are two obvious benefits from Sofi-assistant FLFM. First, Sofi can effectively improve the resolution of each perspective image, so the lateral resolution of the final SOFFLFM image reconstructed by the subsequent deconvolution based on these perspective images will theoretically improve by $\sqrt{n}$ times, compared with that of original FLFM. Furthermore, the improvement of the resolution of the perspective images will also improves the axial resolution of SOFFLFM by $\sqrt{n}$ times theoretically. Second, since the cumulant of gaussian background noise is equal to zero after high-order cumulants processing, the influence of gaussian background noise on original FLFM can be automatically suppressed in SOFFLFM, thus making reconstruction through 3D deconvolution more robust, and reducing artifacts.

## 3. Simulation of SOFFLFM

The SOFFLFM system for the following simulations is similar as what is used in High-resolution Fourier light-field microscopy(HR-FLFM).[8] Three regular hexagonal micro-lenses are used to achieve the minimum segmentation of the full Fourier aperture. According to the calculation in Ref 6, with the parameter settings shown in Table 1, the lateral ($R_{xy}$) and axial ($R_z$) resolution of such an original FLFM system are expected to be 0.508 μm and 0.741 μm respectively, and its depth of focus (DOF) is expected to be 3.78 μm.

**Table 1. System Parameters for Simulation**

| Parameters | σ | λ | NA | M | $f_{TL}$ | $f_{FL}$ | $f_{MLA}$ | $d_{MLA}$ | $P_{sensor}$ |
|---|---|---|---|---|---|---|---|---|---|
| Value | 0.5 | 0.6 μm | 1.45 | 100× | 200 mm | 55 mm | 35 mm | 650 μm | 6.5 μm |

σ : On-state ratio; λ : Emission wavelength; NA : Numerical aperture of objective lens; M : Magnification of objective lens; $f_{TL}$ : Focal length of tube lens; $f_{FL}$ : Focal length of fourier lens; $f_{MLA}$ : Focal length of MLA; $d_{MLA}$ : Pitch of MLA; $P_{sensor}$ : Pixel size.

Although SOFFLFM and HR-FLFM could share the same schematic design, they are totally different in imaging strategy. In SOFFLFM, samples must be labelled with proper fluorophores applicable to Sofi, and a time series of raw images are collected for the consequent cumulants calculation.[13] Referring to the theoretical framework of FLFM,[6] we simulate the 3D PSF of the FLFM system, shown in Fig. 1(b). Each raw image in the time series of SOFFLFM are obtained by convolving the 3D PSF and the distribution of emitters whose fluctuations are modeled following Bernoulli processes. To make the simulations more practical, all the raw images are processed by applying Poisson noise, and adding Gaussian noise whose expected value and variance are set to 0 and 0.01 respectively.

In SOFLFM, raw images are processed as follows. First, the time series of raw images are processed following the same procedure used in Sofi,[14] i.e., calculating the $n^{th}$-order cumulants to generate an image, i.e. $n^{th}$-order SOFFLFM raw image, with improved resolution. Then, Richardson-Lucy iterative deconvolution is applied to this SOFFLFM raw image with corresponding PSF which is actually the $n^{th}$-order cumulative PSF, i.e., the $n^{th}$ power of the original PSF shown in Fig. 1(b). As two examples, the 2nd and the 4th order cumulative PSFs are shown in Fig. 1(c) and (d) respectively. In the following simulations, if there is no specific mentation, the raw images collected for SOFFLFM are sequences consisting of 100 frames, and the number of iterations in deconvolution is 20.

To test the performance of SOFFLFM, a series of simulation and analysis were carried out. First, to verify the improvement in lateral resolution, four pairs of emitters spacing 0.2, 0.3, 0.4, and 0.5 μm were assumed at the same depth, z = 0 μm. According to the parameter settings in Table 1, the effective pixel size is 0.1 μm. Reconstructed images by means of the original FLFM and the SOFFLFM are shown in Fig. 2. FLFM can distinguish two emitters spacing 0.5 μm, which is consistent with the theoretical prediction. For the two emitters separated by 0.4 μm, they cannot be distinguished by FLFM, but obviously distinguishable by 2nd-order SOFFLFM. As to the 4th-order SOFFLFM, two emitters spacing 0.3 μm can easily be distinguished, and two emitters spacing even 0.2 μm could also be distinguished. Compared with the lateral resolution of the FLFM, that of 2nd and 4th order SOFFLFM improved to 1.25 times and more than 2 times respectively.

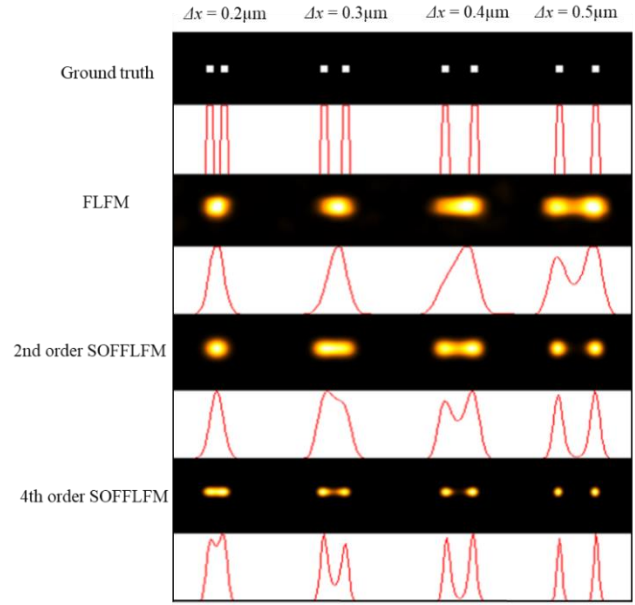

Fig. 2. Comparison of FLFM, 2nd and 4th order SOFFLFM in resolving pairs of emitters with different transversal separations. From top to bottom: the ground-truth, reconstructed results of FLFM, 2nd and 4th order SOFFLFM. Cross-sectional profiles across the centers of each image are attached below correspondingly.

Next, to verify the improved axial resolution, six pairs of emitters were set to different axial separations, from 0.4 μm to 0.6 μm. Reconstructed images by means of the original FLFM and the SOFFLFM are shown in Fig. 3. Similar and solid improvements in axial resolution are observed. FLFM is just able to distinguish emitters spacing 0.9 μm, while 2nd and 4th order SOFFLFM are able to distinguish emitter pairs spacing 0.5 μm and 0.4 μm respectively. Compared with the axial resolution of FLFM, that of 2nd and 4th order SOFFLFM improved to 1.60 times and more than 2.25 times, respectively.

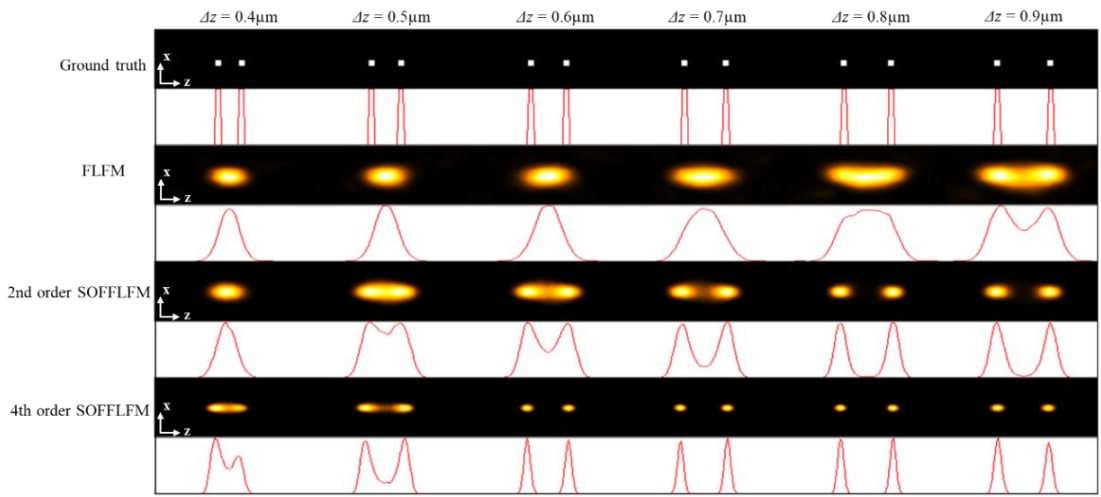

Fig. 3. Comparison of FLFM, $2^{nd}$ and $4^{th}$ order SOFFLFM in resolving pairs of emitters with different axial separations. From top to bottom: the ground-truth, reconstructed images of FLFM, $2^{nd}$ and $4^{th}$ order SOFFLFM. Cross-sectional profiles across the centers of each image are attached below correspondingly.

Then, a sample was simulated, which consists of emitters randomly distributed at four fixed layers with interval of 0.5 μm, as is shown in Fig. 4(e). Raw image collected by FLFM and the corresponding reconstructed FLFM image are shown in Fig. 4(b). and Fig. 4(f). In SOFFLFM, a series of raw images (Fig. 4(a)) are collected, then deal with $2^{nd}$ and $4^{th}$ Sofi algorithm (Fig. 4(c) and Fig. 4(d)), and 3D deconvolved to reconstruct the 3D structure of the sample which are presented as projections on planes x-y, y-z and x-z (Fig. 4(g) and Fig. 4(h)). Compared with the perspective image of FLFM (Fig. 4(b)), that of $2^{nd}$ and $4^{th}$ SOFFLFM (Fig. 4(c) and Fig. 4(d)) show a progressively improved resolution and quite high signal-to-noise ratio (SNR).

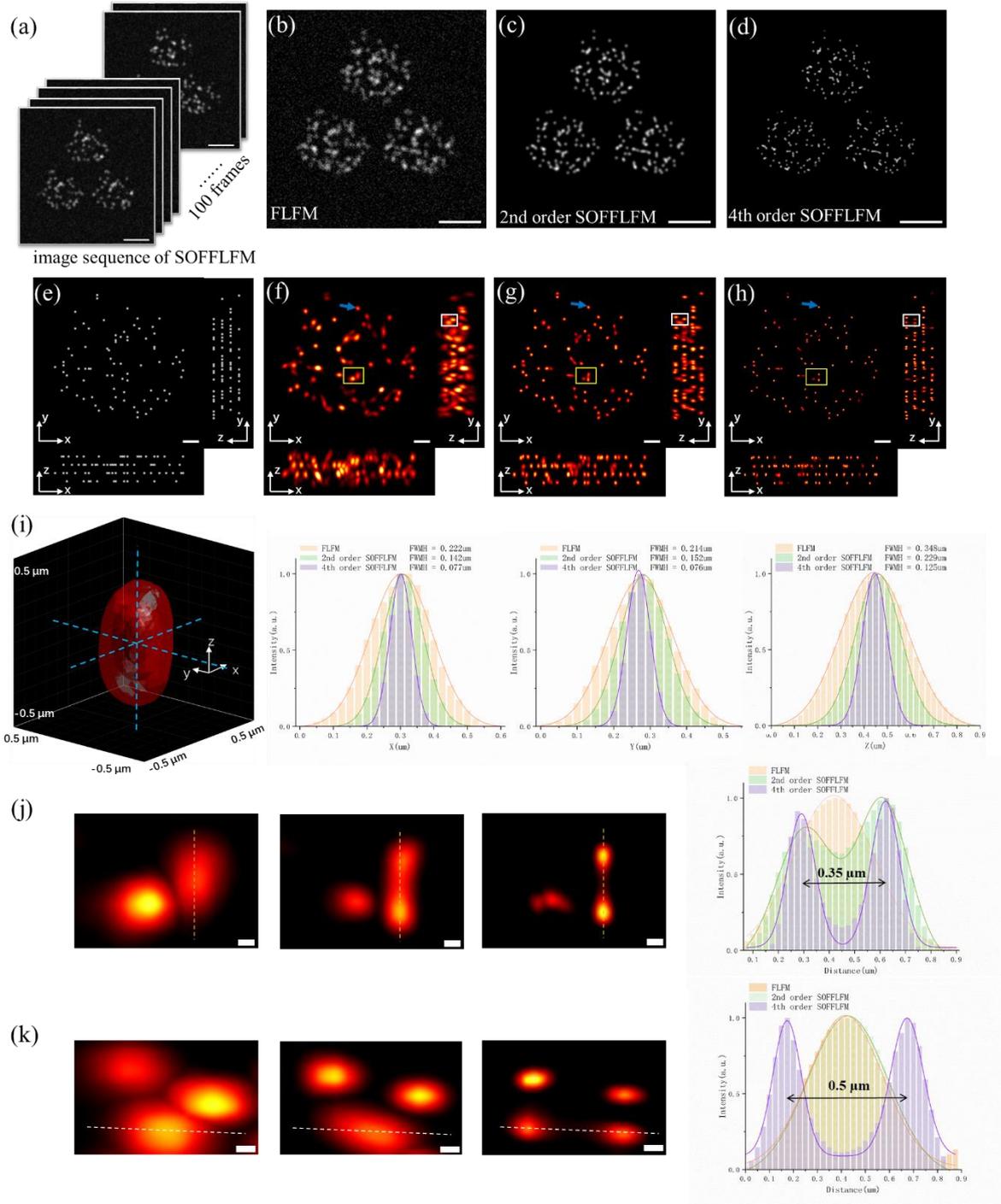

Fig. 4. Raw image of discrete emitters with FLFM (b). (a) A sequence of 100 frames of raw images with SOFFLFM, and their reconstructed raw images of 2$^{nd}$ (c) and 4$^{th}$ (d) order SOFFLFM. (e) Maximum-intensity projection x-y images and the corresponding inset x-z, y-z views of the ground truth volumes. (f-h) Maximum-intensity projection x-y images and the corresponding inset x-z, y-z views of the reconstructed volumes using FLFM(f), 2$^{nd}$ (g) and 4$^{th}$ (h) order SOFFLFM. (i) Left panel, 3D view of a sparse distribution of reconstructed emitter indicated with a blue arrow in (g-f) using the MATLAB function 'isosurface'. Starting from the outer-most isosurface, the corresponding reconstruction results are FLFM, 2$^{nd}$ and 4$^{th}$ order SOFFLFM in turn. Right panel, cross-sectional profiles along the blue dashed lines in x, y, z directions across the center of the emitter, exhibiting FWHM values of 0.222 μm (x), 0.214 μm (y), and 0.348 μm (z)

in FLFM, FWHM values of 0.142 μm (x), 0.152 μm (y), and 0.229 μm (z) in 2$^{th}$-order SOFFLFM, and FWHM values of 0.077 μm (x), 0.076 μm (y), and 0.125 μm (z) in 4$^{th}$-order SOFFLFM, respectively. (j) From left to right are zoomed-in images in x-y of the yellow boxed region in (f-h) and cross-sectional profiles along the yellow dashed lines within these zoom areas. (k) From left to right are zoomed-in images in y-z of the white boxed region in (f-h) and cross-sectional profiles along the white dashed lines within these zoom areas. Scale bar: 10 μm (a, b, c, d), 1 μm (e, f, g, h), 0.1 μm (j, k).

Now, we would like to compare the effect of adjacent emitters on the appearance of a certain emitter in the reconstructed images. Firstly, a sparsely distributed spot (indicated with a blue arrow) from a single emitter was selected for analysis (Fig. 4(i)). In the FLFM image, its full width at half maximum (FWHM) along x, y and z axis are 0.222 μm (x), 0.214 μm (y), and 0.348 μm (z) respectively. In the 2$^{nd}$ order SOFFLFM image, they are 0.142 μm (x), 0.152 μm (y), and 0.229 μm (z), which means an improvement on the lateral resolution by 1.41 times and the axial resolution by 1.52 times compared with that in FLFM. In the 4$^{th}$ order SOFFLFM image, the FWHMs are even squeezed to 0.077 μm (x), 0.076 μm (y) and 0.125 μm (z), increased by 2.82 times and 2.78 times compared with that in FLFM. According to the results above, in FLFM image, an emitter should be distinguishable with its neighbor emitter which is more than 0.222 μm (x or y direction) or 0.348 μm (axial direction) away. However, for the two emitters with even 0.350 μm apart in lateral directions, which are highlighted with yellow rectangles, they could not be resolved (Fig. 4(j)). Two other emitters (highlighted with white rectangles) with 0.500 μm apart in axial direction could not be resolved either (Fig. 4(k)). It seems like the discrete emitters shows better resolution than the non-discrete emitters. Since deconvolution is a standard step for the final reconstructed image, such conclusion is reasonable. And it is another advantage of SOFFLFM, because two emitters which could not distinguished in raw image of FLFM might be resolved in 2$^{nd}$ or 4$^{th}$ order SOFFLFM raw image. That also explain that, compared with the resolution of FLFM, why the resolution improvement of 2$^{nd}$ and 4$^{th}$ order SOFFLFM are more than $\sqrt{2}$ and 2 times (Fig.2 and Fig.3).

Lastly, we use a diverging star-shaped structure as a sample (Fig. 5(a)) to test the influence of different numbers of raw images on the performance of SOFFLFM. Figure. 5(c)(e) and Figure. 5(d)(f) are reconstructed from 100 and 400 frames respectively. Two adjacent lines which could not be resolved in FLFM image (Fig. 5(b)) can be recognized as two lines in 2$^{nd}$-order SOFFLFM images (Fig. 5(c)(d)) and 4$^{th}$-order SOFFLFM images (Fig. 5(e)(f)). However, when the number of frames is small, some information might be lost, especially in high-order SOFFLFM, which can be ascribed to the statistical error of cumulants.

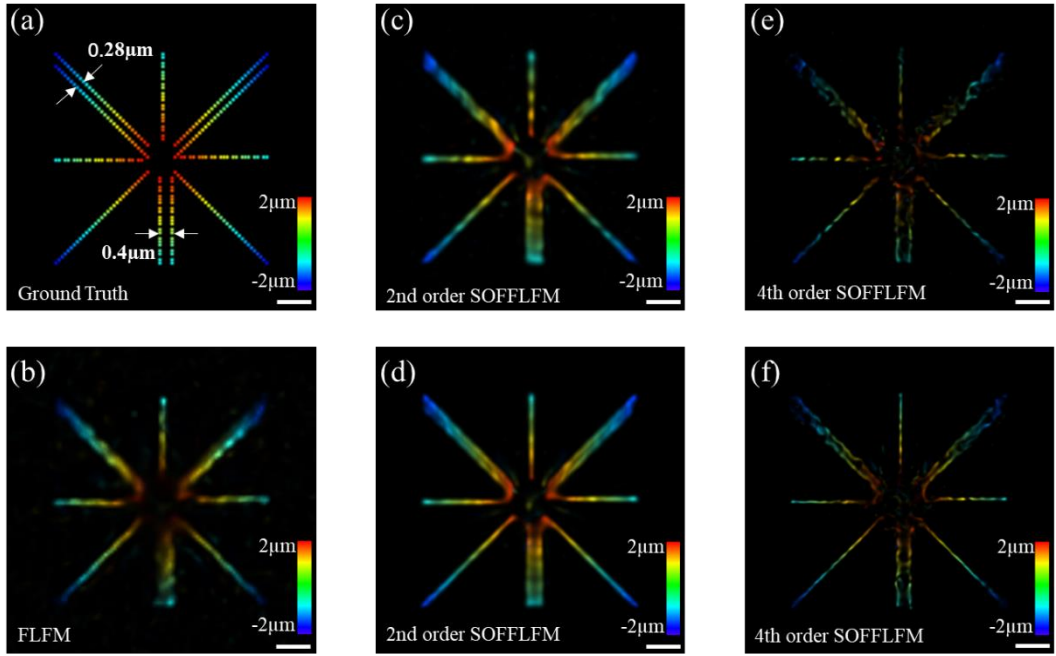

Fig. 5. (a) 3D ground truth. (b) 3D reconstructed result of FLFM. (c), (e) 3D reconstructed results of 2$^{nd}$ (c) and 4$^{th}$(e) order SOFFLFM calculated from the cumulate of 100 frames of image sequence. (d), (f) 3D reconstructed results of 2$^{nd}$ (d) and 4$^{th}$(f) order SOFFLFM calculated from the cumulate of 400 frames of image sequence. The depth information across a 4 μm range in all of the subgraphs is color-coded according to the color scale bar. Scale bar: 1 μm.

## 4. Conclusion

In this paper, we propose an approach to enhance the resolution by introducing high-order cumulants analysis into FLFM. Performances of 2$^{nd}$-order SOFFLFM and 4$^{th}$-order SOFFLFM are tested by simulations. Results demonstrated that more than $\sqrt{2}$ and 2 improvements in resolution in all three dimensions, compared with that in FLFM. Besides, SOFFLFM improve resolution by calculating the cumulants of image sequences collected by normal FLFM systems, which can be directly applied to most FLFM systems without any changes in hardware. In the future, we anticipate SOFFLFM to be an important tool in 3D super-resolution application scenarios.

**Acknowledgments**

This work was supported by The National Natural Science Foundation of China (Grant Nos. 11774242, 61605127, 61975131, 62175166, 61335001), the Shenzhen Science and Technology Planning Project (Grant No. JCYJ20210324094200001, JCYJ20200109105411133)

**Conflicts of Interest**

The authors declare that there are no conflicts of interest relevant to this article.